\documentclass[12pt]{iopart}
\usepackage{graphicx}
\usepackage{color}

\usepackage{cite}
\begin{document}

\title{Bunching of temporal cavity solitons via forward Brillouin scattering}

\author{Miro Erkintalo, Kathy Luo, Jae K. Jang, St\'ephane Coen, and Stuart G. Murdoch}

\address{The Dodd-Walls Centre for Photonic and Quantum Technologies, and Physics Department, The University of Auckland, Private
Bag
92019, Auckland 1142, New Zealand}
\ead{m.erkintalo@auckland.ac.nz}
\vspace{10pt}
\begin{indented}
\item[]August 2015
\end{indented}

\begin{abstract}
  We report on the experimental observation of bunching dynamics with temporal cavity solitons in a
  continuously-driven passive fibre resonator. Specifically, we excite a large number of ultrafast cavity solitons
  with random temporal separations, and observe in real time how the initially random sequence self-organizes into
  regularly-spaced aggregates. To explain our experimental observations, we develop a simple theoretical model that
  allows long-range acoustically-induced interactions between a large number of temporal cavity solitons to be
  simulated. Significantly, results from our simulations are in excellent agreement with our experimental
  observations, strongly suggesting that the soliton bunching dynamics arise from forward Brillouin scattering. In
  addition to confirming prior theoretical analyses and unveiling a new cavity soliton self-organization
  phenomenon, our findings elucidate the manner in which sound interacts with large ensembles of ultrafast pulses
  of light.
\end{abstract}

%
%
%
%
%

\section{Introduction}

Temporal cavity solitons (CSs) are pulses of light that can persist in externally, coherently-driven passive
nonlinear optical resonators~\cite{wabnitz_suppression_1993, leo_temporal_2010}. They are genuine solitons in that
their shape does not evolve upon propagation: temporal broadening induced by chromatic dispersion is balanced by an
optical nonlinearity~\cite{kivshar_optical_2003, agrawal_nonlinear_2006}. In addition, CSs have the ability to
continuously extract energy from the coherent field driving the resonator so as to balance the power losses they
suffer at each cavity roundtrip. This double balancing act makes CSs unique attracting states, and allows them to
circulate indefinitely despite the absence of an amplifier or saturable absorber in the resonator. More generally,
temporal CSs belong to the broader class of localized dissipative structures or dissipative
solitons~\cite{nicolis_self-organization_1977, akhmediev_dissipative_2008, purwins_dissipative_2010}.

Because of the presence of the coherent driving beam, CSs are superimposed and phase-locked on a homogeneous
background field filling the entire resonator~\cite{firth_cavity_2002}. Consequently, they do not possess
phase-rotation symmetry. For that reason, temporal CSs are fundamentally different from pulses in mode-locked
lasers~\cite{firth_temporal_2010}. For example, for the exact same system parameters, a passive driven nonlinear
resonator can sustain at once an arbitrary number of CSs at arbitrary temporal positions. In other words, there are
many different co-existing solutions that the intracavity field can assume and the system exhibits massive
multi-stability (see also~\cite{marconi_how_2014}). Moreover, each of the CSs can be individually addressed, which
means that they can be turned on~\cite{leo_temporal_2010} and off~\cite{jang_writing_2015} and even temporally
shifted with respect to each other~\cite{jang_temporal_2015} without affecting adjacent pulses.

In terms of their identifying characteristics and dynamics, temporal CSs are similar to their spatial counterparts
--- self-localized beams of light that persist in coherently-driven \emph{diffractive} nonlinear
cavities~\cite{ackemann_chapter_2009, barland_cavity_2002}. In particular, both spatial and temporal CSs obey the
paradigmatic mean-field Lugiato-Lefever equation (LLE)~\cite{lugiato_spatial_1987}. But whilst spatial CSs have been
extensively studied for more than two decades~\cite{mcdonald_spatial_1990, tlidi_localized_1994,
firth_optical_1996-1, barland_cavity_2002, lugiato_introduction_2003}, research into temporal CSs only started
in~2010, when they were first observed experimentally by Leo et al using an optical fibre ring
resonator~\cite{leo_temporal_2010}. Due to their unique characteristics, temporal CSs were identified as ideal
candidates for bits in all-optical buffers, which stimulated many subsequent studies using similar fibre cavity
designs~\cite{leo_dynamics_2013, jang_ultraweak_2013, jang_observation_2014, jang_temporal_2015}. In addition to
macroscopic fibre cavities, temporal CSs have also recently attracted great interest in the context of microscopic
Kerr resonators. In particular, it has been shown both theoretically~\cite{coen_modeling_2013, coen_universal_2013,
chembo_spatiotemporal_2013, erkintalo_coherence_2014, godey_stability_2014} and
experimentally~\cite{herr_temporal_2014, brasch_photonic_2014} that temporal CSs are intimately linked to the
formation of broadband Kerr frequency combs that have been observed in such devices~\cite{delhaye_optical_2007,
kippenberg_microresonator-based_2011, moss_new_2013}.

Of course, a defining trait of solitons is concerned with the way they interact with each other
\cite{zabusky_interaction_1965, stegeman_optical_1999, rotschild_long-range_2006}, and CSs are no exceptions.
Theoretical and experimental studies have revealed that CSs are connected to the surrounding background field through
oscillatory tails and that adjacent CSs interact when their tails overlap and/or
lock~\cite{barashenkov_bifurcation_1998, schapers_interaction_2000, tlidi_interaction_2003, bodeker_measuring_2004,
parra-rivas_dynamics_2014}. These interactions can induce rich dynamics in their own right, including the formation
of bound states~\cite{schapers_interaction_2000, jang_controlled_2015}, but are short range due to the exponentially
decaying nature of the oscillatory tails. Experiments with temporal CSs in fibre-based cavities have however also
revealed extremely long range interactions between solitons separated by hundreds of characteristic
widths~\cite{jang_ultraweak_2013}. These were found to be mediated by electrostriction \cite{boyd_nonlinear_2008},
which causes temporal CSs to excite transverse acoustic waves in the fibre core and cladding. The acoustic waves give
rise to refractive index changes through the acousto-optic effect, and long-range interactions ensue when a trailing
CS overlaps with the perturbation created by a leading one~\cite{jang_ultraweak_2013}.

Long before temporal CSs were even observed, electrostriction-induced interactions were studied in the context of
optical-fibre telecommunication systems~\cite{smith_experimental_1989, dianov_long-range_1992,
jaouen_transverse_2001}, as well as in passively mode-locked fibre lasers~\cite{grudinin_passive_1993,
pilipetskii_acoustic_1995, grudinin_passive_1997}. In particular, Pilipetskii et al have numerically demonstrated
that acoustic effects could be responsible for the bunching of pulses in fibre
lasers~\cite{pilipetskii_acoustic_1995}. Although experimental observations of pulse bunching abound in the ultrafast
fibre laser literature~\cite{grudinin_passive_1993, tang_bound-soliton_2002, zhao_passive_2009, chouli_rains_2009},
quantitative comparisons with the theory of acoustic interactions are hindered by the many competing effects that
influence pulse dynamics in such lasers, including saturable absorption and gain depletion and
recovery~\cite{grudinin_passive_1997, kutz_stabilized_1998, korobko_long-range_2015}. Continuously-driven passive
fibre cavities are void of these complications, and acoustic interactions of a pair of temporal CSs have been
successfully modelled quantitatively~\cite{jang_ultraweak_2013}. Moreover, because temporal CSs are phase-locked to
the cavity driving beam, the acoustic interactions they experience are orders of magnitude weaker than in other
systems~\cite{jang_ultraweak_2013}. The pertinent dynamics can therefore be easily monitored in real time. Temporal
CSs in coherently-driven passive fibre cavities thus appear as the ideal platform to explore
electrostriction-mediated pulse interaction effects. So far, however, experiments with temporal CSs have only been
performed with a small number of co-existing solitons~\cite{jang_ultraweak_2013, jang_temporal_2015}. Accordingly, no
pulse bunching effects have yet been observed.

In this Article, we experimentally and theoretically investigate the acoustic interactions of a very large number of
temporal CSs. Specifically, we excite a large number of randomly-spaced temporal CSs in a continuously-driven passive
fibre cavity (hence based on a simple Kerr nonlinearity), and we examine their interaction dynamics in real time. We
find that the initially random sequence of pulses self-organizes into regular bunches whose spacing agrees very well
with the frequency of the acoustic modes that interact most efficiently with light in the fibre core. To
quantitatively show that the bunching behaviour originates from acoustic effects, we develop a simple model that
allows the full dynamical evolution to be simulated. Very good agreement is observed between simulations and
experiments.

\section{Experiment}

\subsection{Experimental setup}

\begin{figure}[t]
  \includegraphics[width=1\textwidth, clip = true]{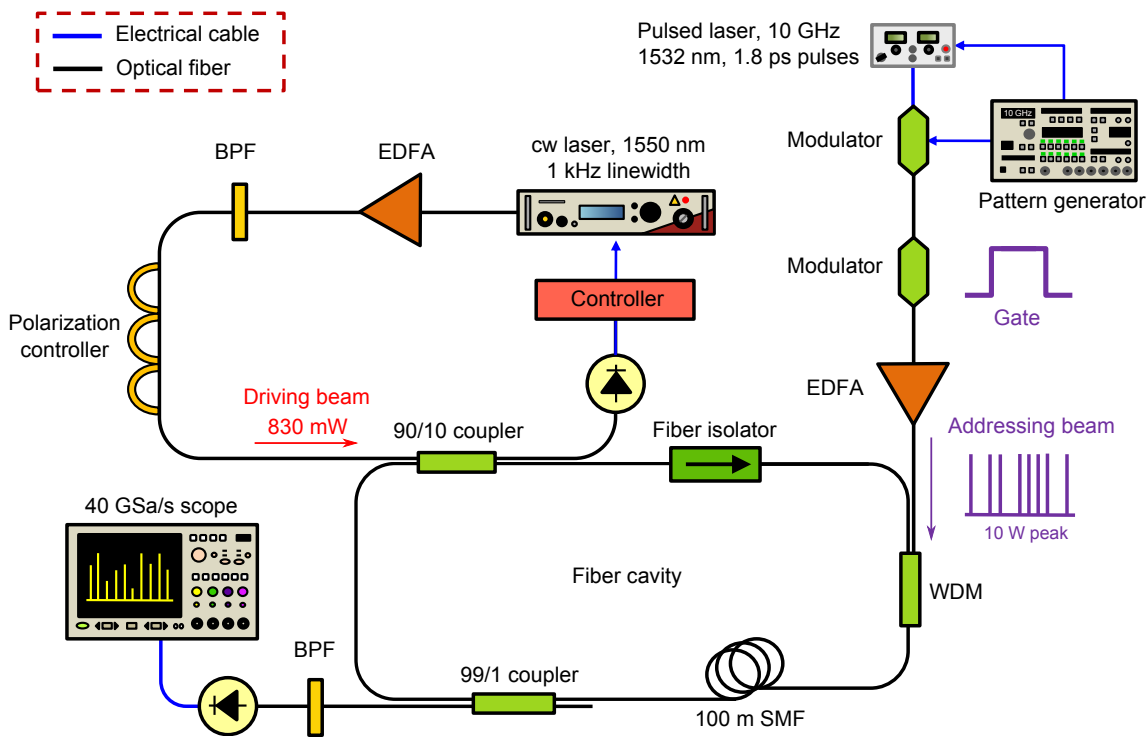}
  \caption{Experimental setup.}
  \label{setup}
\end{figure}

Our experimental setup is similar to the one used in~\cite{jang_ultraweak_2013}, and is schematically illustrated in
Fig.~\ref{setup}. The passive all-fibre cavity is 100-m long and constructed of standard telecommunications
single-mode optical fibre (SMF28) that is closed on itself with a 90/10 fibre coupler. The cavity incorporates an
optical isolator to prevent depletion of the driving beam by \emph{backward} stimulated Brillouin
scattering~\cite{agrawal_nonlinear_2006}, a wavelength-division multiplexer (WDM) to couple in addressing pulses used
to ``excite'' the  CSs (see below), and a 1\,\% output coupler through which the intracavity CS dynamics can be
monitored. Overall, the cavity has a finesse of $21.5$, corresponding to $29.2$\,\% power losses per roundtrip.

The cavity is coherently driven with an ultra-narrow linewidth ($<1$~kHz) continuous-wave (cw) laser centred at
1550~nm wavelength, that is externally amplified to about 1~W with an erbium-doped fibre amplifier (EDFA). Amplified
spontaneous emission noise is removed using a $0.6$-nm-wide bandpass filter (BPF) centred at 1550~nm before the field
is injected into the cavity via the 90/10 fibre coupler. The light that is reflected off from the cavity is fed to a
servo-controller that actuates the driving laser frequency so as to maintain the reflected power at a set level. In
this way, the frequency of the driving laser follows any changes in the cavity resonances due, e.g., to environmental
perturbations, ensuring that the phase detuning between the driving laser and the cavity is locked. This is an
integral part of our experiment, as temporal CSs rely critically on phase-sensitive interactions with the driving
field. Note that the cavity locking scheme employed here is more robust than that used in the first experimental
observation of Leo et al~\cite{leo_temporal_2010}; in our setup, CSs can routinely be sustained for several minutes
or even hours. This stability is crucial to our study, since the acoustic CS interactions are so weak that very long
measurement times are necessary to observe the full dynamics~\cite{jang_ultraweak_2013}.

To excite temporal CSs, we use the optical addressing technique introduced in~\cite{leo_temporal_2010}. Specifically,
ultrashort pulses from a 10~GHz mode-locked laser with a different wavelength (here, 1532~nm) than that of the
driving field are launched into the cavity through the WDM. They then interact through cross-phase modulation with
the intracavity cw background and each of them excites an independent CS. After one roundtrip, the addressing pulses
exit the cavity through the WDM, and only the temporal CSs persist. The process is controlled by picking pulses from
the mode-locked laser with a sequence of two intensity modulators. The first modulator is driven by a 10~GHz pattern
generator synchronized to the mode-locked laser and selects the pattern of CSs to be excited. The second modulator is
used as a gate to block the mode-locked laser beam after addressing is complete. Once CSs are excited, we monitor
their dynamics in real time by recording the field at the cavity output using a fast $12.5$~GHz photodetector
connected to a 40~GSa/s oscilloscope. Before detection, the output field passes through a narrow bandpass filter
centred at 1551~nm, one nanometer away from the driving wavelength. This removes the cw background component of the
CSs, improving the signal-to-noise ratio of the measurements~\cite{leo_temporal_2010}.

\subsection{Experimental results}

In previous studies, we have examined configurations involving a small number of temporal
CSs~\cite{jang_ultraweak_2013, jang_temporal_2015}. Here, in contrast, we are interested in studying the intracavity
dynamics when a very large number of CSs co-exist. To this end, we start the experiment by exciting a densely-packed
sequence of temporal CSs. This is achieved by programming a random sequence into the pattern generator driving the
first modulator, in essence selecting a corresponding random series of pulses from the mode-locked addressing laser,
while the second modulator is kept open for \emph{several} cavity roundtrip times~$t_\mathrm{R}$. Because the
mode-locked laser repetition rate, the length of the random sequence, and the cavity free-spectral-range
($\mathrm{FSR} = 1/t_\mathrm{R}$) are not commensurate, the resultant temporal CS sequence is to a large extent
random.

\begin{figure}[t]
  \includegraphics[width=1\textwidth, clip = true]{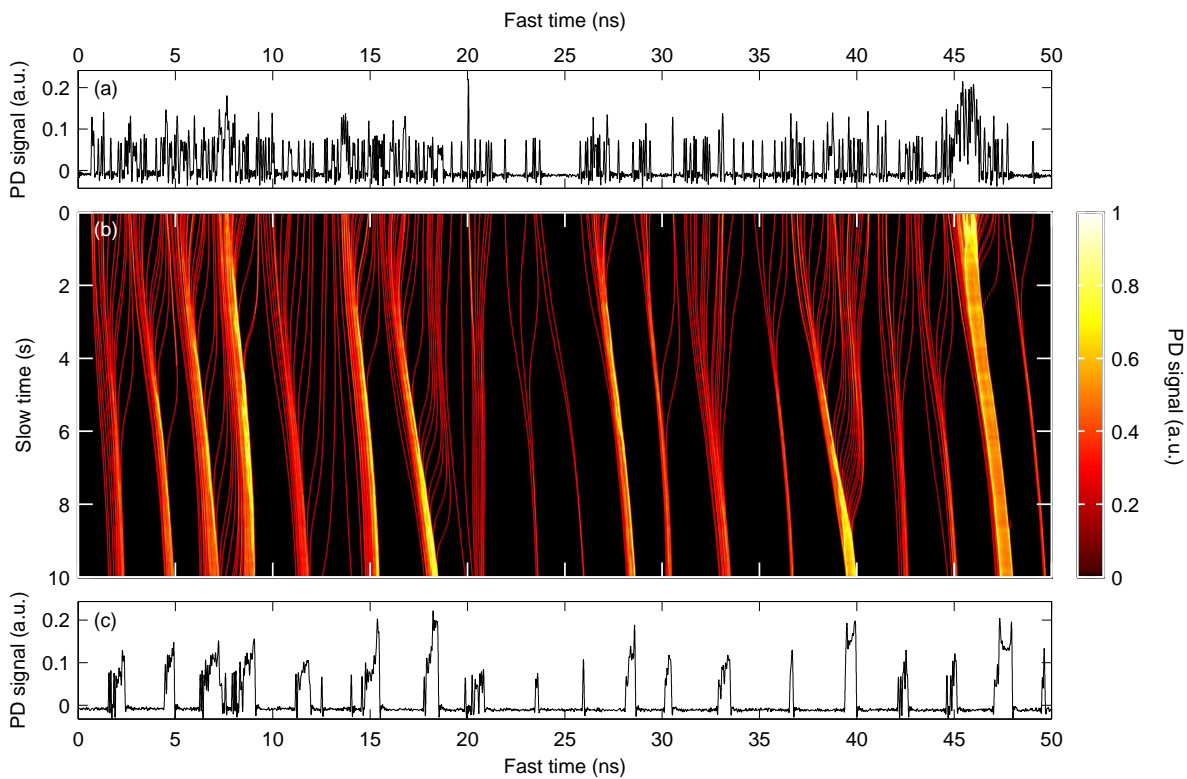}
  \caption{Experimental results. (a, c) Temporal intensity profiles of the intracavity field measured (a) right after
    exciting temporal CSs, and (c) after those CSs have freely interacted for 10~seconds. (b) Density plot
    corresponding to a vertical concatenation of such profiles measured at regular intervals and revealing how the field
    dynamically evolves over time (top to bottom). The top and bottom lines correspond to the profiles shown in
    (a) and (c), respectively. PD: photodiode.}
  \label{expr}
\end{figure}

The curve in Fig.~\ref{expr}(a) illustrates the result of the addressing process. It shows the temporal intensity
profile of the intracavity field recorded by the oscilloscope at the beginning of the experiment, highlighting the
presence in the cavity of a sequence of temporal CSs with essentially random spacing. Note that (i) for clarity we
only show a small 50~ns-long segment of the full 480-ns roundtrip, and (ii) that the electronic bandwidth of our
detectors prevents closely-spaced temporal CSs to be individually resolved. In this context, we remark that, for
given parameters, all temporal CSs have identical characteristics (energy, duration, and peak
power)~\cite{leo_temporal_2010}. The different amplitudes observed in Fig.~\ref{expr}(a) therefore simply represent
bunches that contain different numbers of temporal CSs spaced by less than the detector 80~ps response time.

In Fig.~\ref{expr}(c) we show a similar measurement but taken approximately 10~s after the temporal CSs were excited
and allowed to freely interact. Here we see clearly that the CSs have formed almost regularly spaced aggregates, with
an average separation of about~$2.6$~ns. This bunching behaviour can be more readily appreciated from the false
colour density plot in Fig.~\ref{expr}(b), which maps the measured dynamical evolution of the CS field during the
10~s of free interaction. To form this plot, we have vertically concatenated 100 oscilloscope profiles [like those
shown in Figs~\ref{expr}(a) and (c)] measured at regular intervals (10 frames/s) so as to display how the intracavity
pulse sequence evolves over time (top to bottom). We see clearly how the CSs exhibit complex interaction dynamics,
with individual pulses gradually forming bunches. To the best of our knowledge, this represents the first direct
experimental observation of pulse bunching dynamics in a fibre resonator. We also highlight that, as
in~\cite{jang_ultraweak_2013}, the interactions are exceedingly weak. During the 10~s measurement shown in
Fig.~\ref{expr}, the temporal CSs complete about 20~million roundtrips (corresponding to 2~million kilometres of
propagation length), yet their temporal separations only change by a few nanoseconds.

\section{Theory}

In this Section, we show theoretically that the bunching dynamics observed in the experiment described above can be
quantitatively explained in terms of electrostriction-induced acoustic interactions. We first recount the basic
mechanisms that underpin the interactions, and subsequently develop a simple model that allows the acoustic
interactions of a large number of temporal CSs to be examined. Our approach is adapted from that developed by
Pilipetskii et al to investigate acoustic interactions in passively mode-locked fibre
lasers~\cite{pilipetskii_acoustic_1995}.

\subsection{Acoustic soliton interactions}

Pulses of light travelling in optical fibres can excite, through electrostriction, transverse acoustic waves
propagating (nearly) orthogonally to the fibre axis~\cite{jaouen_transverse_2001, boyd_nonlinear_2008}, giving rise
to refractive index perturbations that are left behind in the wake of the excitation pulses. The physical mechanism
coincides with guided acoustic wave Brillouin scattering (also referred to as forward Brillouin scattering) that was
first studied by Shelby et al in the context of cw fields~\cite{shelby_resolved_1985, shelby_guided_1985}. Dianov et
al~\cite{dianov_long-range_1992} were the first who suggested that this mechanism could explain long-range interpulse
interactions previously observed in optical fibres~\cite{smith_experimental_1989}.

\begin{figure}[b]
  \includegraphics[width=1\textwidth, clip = true]{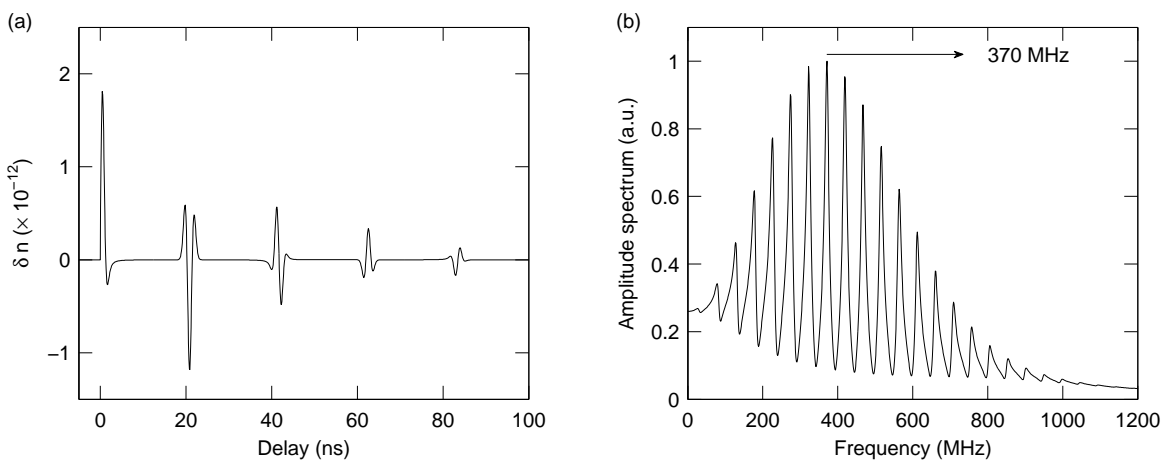}
  \caption{Acoustic-induced refractive index perturbation created by a temporal CS for the parameters of our experiment.
    (a) Time-domain impulse response and (b) its amplitude spectrum. The spectral maximum occurs at a frequency of
    370~MHz, typical of single-mode silica optical fibres.}
  \label{acoustic}
\end{figure}

Figure~\ref{acoustic}(a) is a plot of the temporal impulse response of the effective refractive index perturbation,
$\delta n(\tau)$, generated through this process in the fibre core. The response was calculated following the
approach of Dianov et al~\cite{dianov_long-range_1992} and using parameters (in particular the CS energy)
corresponding to our experiment (see Ref.~\cite{jang_ultraweak_2013} for details). The perturbation is fairly weak
but extends over tens of nanoseconds. The overall shape of the response is dominated by 1--3~ns wide spikes that are
separated by about 21~ns, arising from successive acoustic reflections on the fibre cladding-jacket boundary. We must
note that the temporal CSs in our experiment have a $\sim 3$~ps duration. The impulse response shown
in Fig.~\ref{acoustic}(a) is thus a fair representation of the refractive index perturbation induced by acoustic
waves generated through electrostriction by an isolated temporal CS.

The refractive index perturbation shown in Fig.~\ref{acoustic}(a) is continuously generated by a temporal CS as it
propagates down the fibre at the speed of light, and exists as a spatially extended tail behind it. Due to its
time-dependence, it can affect the group velocity of a trailing temporal CS, thus giving rise to long-range
interactions. Specifically, if a temporal CS overlaps with a portion of the $\delta n(\tau)$ perturbation that has a
negative (positive) gradient, the CS will speed up (slow down), leading to a time-domain drift towards the maxima of
the refractive index change induced by the CSs leading it. For the case of two temporal CSs, the perturbation is
simply given by the impulse response shown in Fig.~\ref{acoustic}(a). Accordingly, the trailing CS will increase or
decrease its separation from the leading one until it coincides with one of the maxima of the response shown in
Fig.~\ref{acoustic}(a)~\cite{jang_ultraweak_2013}.

When more than two temporal CSs are involved, the dynamical evolution of a particular one is affected by the
superposition of refractive index perturbations induced by all the temporal CSs leading it. In general, this
superposition can assume a very complex temporal profile. Pilipetskii et al have however numerically shown, in the
context of passively mode-locked fibre lasers~\cite{pilipetskii_acoustic_1995}, that a large sequence of light pulses
may spontaneously form bunches whose separations correspond to the acoustic frequency that interacts the most
efficiently with light. To gauge whether this hypothesis is related to our experimental observations, we plot in
Fig.~\ref{acoustic}(b) the absolute value of the electrostrictive frequency response in our system, i.e.,
$|\mathcal{F}\left[\delta n(\tau)\right]|$, where $\mathcal{F}[\cdot]$ denotes Fourier transformation and $\delta
n(\tau)$ is the impulse response shown in Fig.~\ref{acoustic}(a). As can be seen, maximum spectral amplitude is
reached at a frequency of 370~MHz, and the corresponding $2.7$~ns period is in very good agreement with the $2.6$~ns
bunch spacing observed in the experimental results of Fig.~\ref{expr}. This strongly suggests that the observed
bunching dynamics is indeed due to acoustic interactions of the very large number of temporal CSs.

\subsection{Simulation model}

To establish quantitatively that electrostriction-induced interactions can explain our experimental observations, we
have performed numerical simulations of the underlying dynamics. In Ref.~\cite{jang_ultraweak_2013}, a nonlinear
partial differential equation was derived that was shown to accurately model the dynamics of temporal CSs and their
acoustic interactions. Unfortunately, direct brute force simulations of that model is not computationally feasible
here due to the very large number of temporal CSs and the extremely different timescales involved. We instead develop
and use a simplified model that de-couples the soliton physics from the acoustic
effects~\cite{pilipetskii_acoustic_1995}. Specifically, we represent the entire CS sequence using only the temporal
positions~$\tau_i$ of the individual solitons ($i=1$, 2, 3,~$\ldots$), and we examine how those positions evolve over
time under the influence of acoustic waves generated by the corresponding CSs. To this end, we first need to
quantitatively establish how the velocity of a CS is modified by a given refractive index perturbation. In this
context, we note that temporal CSs in passive cavities react very differently to perturbations than pulses in
mode-locked fibre lasers, and we therefore cannot simply use the approach of~\cite{pilipetskii_acoustic_1995}.

We start by considering the full partial-differential model of a Kerr cavity (a generalized LLE) that takes acoustic
refractive index perturbations into account~\cite{jang_ultraweak_2013}. Assuming the CSs act as Dirac-$\delta$
functions in exciting acoustic waves, the evolution of the intracavity field $E(t,\tau)$ can be written in
dimensionless form as~\cite{jang_ultraweak_2013}:
\begin{equation}
\centering
  \frac{\partial E(t,\tau)}{\partial t} = \left[ -1-i\Delta+i|E|^2+i\frac{\partial^2}{\partial\tau^2}\right]E + S +i\nu(\tau)E.
  \label{eq:LL}
\end{equation}
The normalization of this equation is the same as that used in the Supplementary Information of
Ref.~\cite{leo_temporal_2010}. The variable~$t$ corresponds to the \emph{slow} time of the resonator that describes
evolution of the field envelope~$E$ at the scale of a photon lifetime, whilst $\tau$ is a \emph{fast} time describing
the temporal profile of the field envelope. The first five terms on the right-hand side of Eq.~(\ref{eq:LL})
describe, respectively, the total cavity losses, phase detuning of the pump from resonance (with $\Delta$ the
detuning coefficient), Kerr nonlinearity, anomalous group-velocity dispersion, and external driving (with $S$ the
amplitude of the cw driving field).

\begin{figure}[b]
  \includegraphics[width=1\textwidth, clip = true]{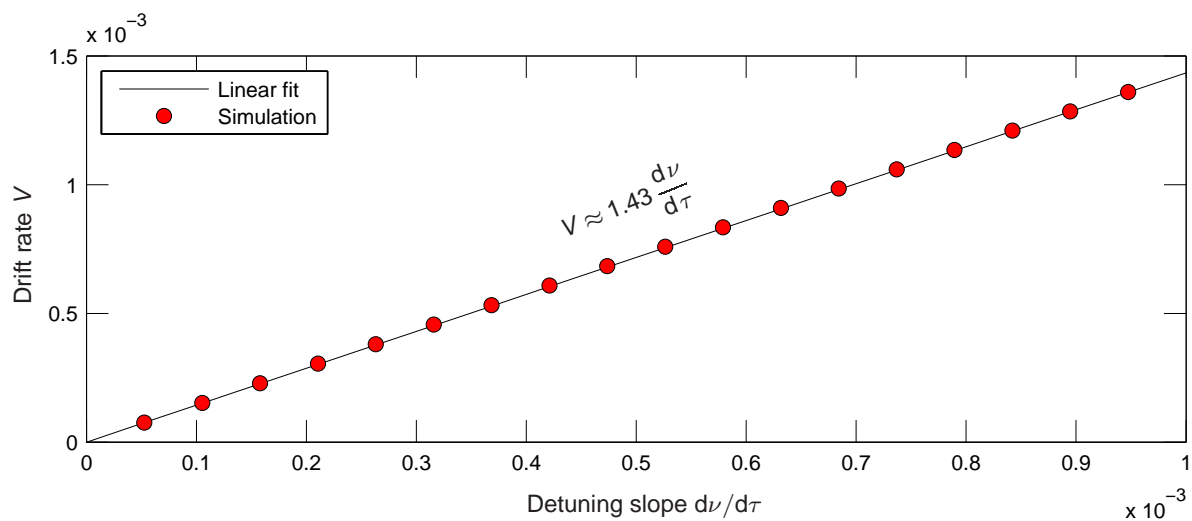}
  \caption{Red circles show numerically simulated CS drift rates for parameters corresponding to our experiments
  ($S^2\approx 2.65$, $\Delta = 2.3$) as a function of the detuning gradient~$A$. The black solid line is a linear fit
  with a slope of~$1.43$.}
  \label{sim1}
\end{figure}

The last term on the right-hand side of Eq.~(\ref{eq:LL}) describes the (normalized) acoustic-induced refractive
index perturbation created by the temporal CSs present in the field~$E(t,\tau)$. As can be seen, it amounts to
introducing a time-dependent perturbation to the cavity detuning~$\Delta$. Earlier studies of spatial CSs have
revealed that detuning perturbations cause CSs to alter their velocities in proportion to the gradient of the
perturbation~\cite{maggipinto_cavity_2000, caboche_cavity-soliton_2009}. To verify this behaviour, and also to find
the proportionality constant for our parameters [see caption of Fig.~\ref{sim1}], we have numerically integrated
Eq.~(\ref{eq:LL}) for a wide variety of different perturbation gradients. Specifically, we ran a set of simulations
with detuning perturbations of the form $\nu(\tau) = A\tau$, where the gradients~$A$ were chosen to have similar
magnitudes to those arising from acoustic effects. For each value of~$A$, we started the simulation with a single
temporal CS centred at $\tau = \tau_i$, and we extracted the rate at which its temporal position drifts: $V =
\mathrm{d}\tau_i/\mathrm{d} t$.

Figure~\ref{sim1} shows results from these simulations. A linear relationship between the CS drift rate~$V$ and the
detuning gradient is evident. For our experimental conditions, we can thus approximate $V =
\mathrm{d}\tau_i/\mathrm{d}t \approx r\,\mathrm{d}\nu/\mathrm{d}\tau|_{\tau_i}$, with the proportionality constant
$r\approx 1.43$. Transforming to dimensional units, we find that the CS temporal positions~$\tau_i$ obey the
following first-order ordinary differential equation
\begin{equation}
  \centering
  \frac{\mathrm{d}\tau_i}{\mathrm{d}t}=r\frac{|\beta_2|L^2\mathcal{F}}{t_\mathrm{R}\lambda_0}
    \left.\frac{\mathrm{d}n_{\mathrm{tot}}}{\mathrm{d}\tau}\right|_{\tau_i}.
  \label{taui}
\end{equation}
Here $\beta_2$ is the fibre group-velocity dispersion coefficient, $L$ is the cavity length, $\mathcal{F}$ is the
cavity finesse, $t_\mathrm{R}$ is the cavity roundtrip-time, and $\lambda_0$ is the wavelength of the driving field.
Finally, $n_{\mathrm{tot}}(\tau)$ corresponds to the total acoustic-induced refractive index perturbation existing in
the cavity. It is given by
\begin{equation}
  n_{\mathrm{tot}}(\tau) = \sum_{j}\delta n(\tau-\tau_j) + \sum_{j}\delta n(t_\mathrm{R}+\tau-\tau_j).
  \label{ntot}
\end{equation}
where $\delta n(\tau)$ is the impulse response introduced above. Given the causal nature of $\delta n(\tau)$, the
first term represents simply the superposition of the index perturbations induced by all CS present before
time~$\tau$. For single-pass propagation through an optical fibre, this term alone would appear. For a fibre cavity,
one must however also take into account the periodic nature of the boundary conditions. Specifically, a temporal CS
completing its $m^{th}$ roundtrip across the cavity may be affected by index perturbations induced by CSs during the
$(m-1)^{th}$ roundtrip. This is accounted for by the second term in Eq.~(\ref{ntot}). Note that our cavity roundtrip
time $t_\mathrm{R} = 480$~ns is much longer than the lifetime of the acoustic waves [see Fig.~\ref{acoustic}(a)], and
therefore only CS present at temporal positions behind time~$\tau$ contribute to this term in practice. For the same
reason, perturbations created more than one roundtrip earlier do not need to be considered.

\subsection{Simulation results}

\begin{figure}[t]
  \includegraphics[width=1\textwidth, clip = true]{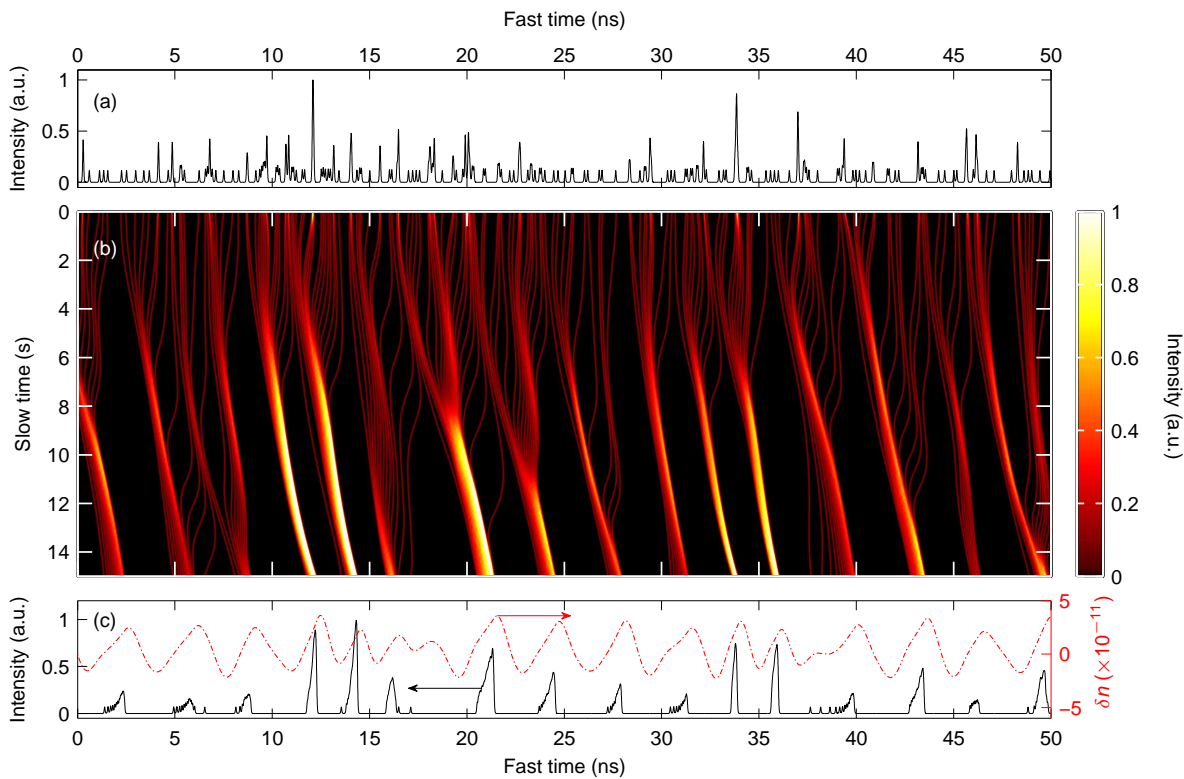}
  \caption{Simulation results. (a, c) Solid lines show the temporal intensity profiles of the CS sequence (a) in
  the beginning of the simulation and (c) after the CSs have freely interacted for 15 seconds. The dash-dotted line in
  (c) is the total refractive index perturbation $n_{\mathrm{tot}}(\tau)$ at the end of the simulation.
  (b) Density plot revealing the full simulated dynamical evolution taking place between (a) and (c) as in
  Fig.~\ref{expr}. The parameters used in our simulations are similar to experimental values:
  $\beta_2 = -21.4~\mathrm{ps^2/km}$; $\mathcal{F} = 21.5$; $L = 100$~m; $t_\mathrm{R} = 0.48~\mathrm{\mu s}$; $\lambda_0 =
  1550$~nm.}
  \label{sim2}
\end{figure}

Equations~\ref{taui} and \ref{ntot} make it possible to efficiently simulate the acoustic interactions of an
arbitrary sequence of temporal CSs. Figure~\ref{sim2} shows results from numerical integration using parameters
corresponding to our experiment above [and listed in the caption of Fig.~\ref{sim2}]. Since extracting the precise
random CS sequence that was excited in our experiment is difficult, we assume here that the cavity initially contains
2000 CSs whose temporal positions follow a uniform random distribution.

Figures~\ref{sim2}(a) and (c) show 50-ns-long snapshots of the temporal intensity profiles of the CS sequences at the
beginning and end of the simulation, respectively, while Fig.~\ref{sim2}(b) reveals the full dynamics over 15~s,
using the same representation as for the experimental data in Fig.~\ref{expr}. To facilitate visualization, and to
mimic the temporal resolution of the oscilloscope, each CS is represented as a $\mathrm{sech}$ profile with 80~ps
full-width at half maximum. The simulated CS interaction dynamics is clearly in excellent agreement with the
experimental observations. In particular, we see that the initial random sequence [Fig.~\ref{sim2}(a)] self-organizes
into regularly-spaced bunches [Fig.~\ref{sim2}(c)]. We note that the about 3~ns average bunch spacing observed at the
output of our simulations is somewhat larger than the $2.6$~ns experimental figure. This discrepancy is attributed to
an imperfect knowledge of the acoustic impulse response that was already identified in \cite{jang_ultraweak_2013}.
Compounded by further uncertainties in the initial CS configuration and other experimental parameters, this could
also explain the slightly different time-scale over which the bunching takes place in our experiment in comparison to
simulations.

To further confirm our interpretation, we have plotted as a red dash-dotted line in Fig.~\ref{acoustic}(c) the total
refractive index perturbation~$n_{\mathrm{tot}}(\tau)$ at the end of the simulation. This perturbation assumes an
almost sinusoidal shape, and its approximately 3~ns period remarkably matches the simulated bunch spacing. It is also
very clear from the dynamical evolution trajectories [Fig.~\ref{sim2}(b)] that the temporal CS bunches experience an
overall drift towards the perturbation maxima. However, even when continuing the simulations over much longer time
scales, the bunches never reach the maxima, and always stay slightly offset from them [this is already visible in
Fig.~\ref{sim2}(c)]. In that way, the bunches eventually reach a quasi-stationary state in which they all drift with
a non-zero near-constant velocity, chasing the maxima. Of course, the maxima themselves keep shifting in the same
direction, as they are constantly re-formed by the drifting temporal CSs. This probably constitutes a general feature
of this kind of interaction. Moreover, we suspect that the drift of the refractive index perturbation explains why
the solitons in each bunch do not come arbitrarily close to each other --- in our simulations, the CSs stay spaced by
some tens of picoseconds within a bunch. If the refractive index perturbation was stationary, the CSs present within
each bunch would meet at a maximum, where they would merge into one or annihilate~\cite{jang_controlled_2015}. In
that scenario, the bunches would progressively disappear, which is clearly not consistent with our experimental
observations (see Fig.~\ref{expr}).

\section{Conclusions}

To conclude, we have experimentally and theoretically studied the acoustic dynamics of a very large number of
temporal CS in a coherently-driven passive fibre resonator. Our experiment reveals that the CSs exhibit complex
interactions, resulting in the formation of almost regularly-spaced bunches, each made up of multiple CSs. To explain
our observations, we have developed a simple theoretical framework that allows the electrostriction-induced
long-range interactions of arbitrary temporal CS sequences to be simulated. Numerical results are in very good
agreement with experimental observations, confirming that the observed bunching dynamics originate from the
excitation of transverse acoustic waves. In addition to unveiling a new dynamical behaviour of temporal CS ensembles,
our results quantitatively confirm the 1995 theoretical predictions of Pilipetskii et al concerning pulse bunch
formation via electrostriction-induced interactions. We expect our result to greatly expand our understanding of
temporal CSs and their interactions, as well as the manner in which sound interacts with long sequences of ultrafast
pulses of light.

\section*{Acknowledgements}

We acknowledge support from the Finnish Cultural Foundation and the Marsden Fund of The Royal Society of New Zealand.

\section*{References}

\end{document}